\documentclass[conference]{IEEEtran}
\IEEEoverridecommandlockouts
\usepackage{amsmath,graphicx}
\usepackage{color,soul}
\usepackage{arydshln}
\usepackage{mathrsfs}
\usepackage{amssymb}
\usepackage{bbm}
\usepackage{dsfont}
\usepackage{cite}
\usepackage{comment}

\newcommand{\mbf}[1]{\mathbf{#1}}
\newcommand{\bb}[1]{\boldsymbol{#1}}

\usepackage{arydshln}
\usepackage{algorithm,algcompatible,amsmath}
\makeatletter
\renewcommand{\ALG@name}{Table}
\makeatother
\usepackage{epstopdf}
\usepackage{graphicx}
\usepackage{multicol}
\usepackage{cite}
\usepackage{thmtools}
\usepackage{hyperref}

\usepackage{enumitem}

\usepackage{subcaption} 
\usepackage{balance}
\usepackage{tikz}
\usepackage{lipsum}
\def\BibTeX{{\rm B\kern-.05em{\sc i\kern-.025em b}\kern-.08em
    T\kern-.1667em\lower.7ex\hbox{E}\kern-.125emX}}

\newcommand\copyrighttext{%
	\footnotesize 
	Copyright 2018 IEEE. Published in the IEEE 2018 Asilomar Conference on Signals, Systems, and Computers (Asilomar 2018), scheduled for 28-31 October 2018 in Asilomar, Pacific Grove, CA, USA. Personal use of this material is permitted. However, permission to reprint/republish this material for advertising or promotional purposes or for creating new collective works for resale or redistribution to servers or lists, or to reuse any copyrighted component of this work in other works, must be obtained from the IEEE. Contact: Manager, Copyrights and Permissions / IEEE Service Center / 445 Hoes Lane / P.O. Box 1331 / Piscataway, NJ 08855-1331, USA. Telephone: + Intl. 908-562-3966.}
\def\copyrightnotice{%
	\begin{tikzpicture}[remember picture,overlay]
	\node[anchor=south,yshift=0pt] at (current page.south) {\fbox{\parbox{\dimexpr\textwidth-\fboxsep-\fboxrule\relax}{\copyrighttext}}};
	\end{tikzpicture}%
}

\begin{document}

\title{Signal Recovery From 1-Bit Quantized Noisy Samples via Adaptive Thresholding}
\author{Shahin Khobahi$^{\star}$ and Mojtaba Soltanalian\\
	Department of Electrical and Computer Engineering, University of Illinois at Chicago, 
	Chicago, USA 
	\thanks{$^{\star}$ Corresponding author (e-mail: \textit{skhoba2@uic.edu}). This work was supported in part by U.S. National Science Foundation Grants CCF-1704401 and ECCS-1809225.}
}

\maketitle
\copyrightnotice
\begin{abstract}
In this paper, we consider the problem of signal recovery from 1-bit noisy measurements. We present an efficient method to obtain an estimation of the signal of interest when the measurements are corrupted by white or colored noise. To the best of our knowledge, the proposed framework is the pioneer effort in the area of 1-bit sampling and signal recovery  in providing a unified framework to deal with the presence of noise with an arbitrary covariance matrix including that of the colored noise. The proposed method is based on a constrained quadratic program (CQP) formulation utilizing an adaptive quantization thresholding approach, that further enables us to accurately recover the signal of interest from its 1-bit noisy measurements. In addition, due to the adaptive nature of the proposed method, it can recover both fixed and time-varying parameters from their quantized 1-bit samples. 
\end{abstract}

\section{Introduction}
Wireless sensor networks (WSNs) present significant potential for usage in spatially wide-scale detection and estimation. However, there exist several practical constraints such as a low power consumption, a low cost of manufacturing, and more importantly, limited computational and transmission capabilities that must be addressed in implementing such networks \cite{SM}. As a result, it is highly desirable to develop distributed estimation frameworks with which the nodes can reliably communicate with the fusion center (FC) while satisfying the power and computational constraints. Furthermore, quantization of signal of interest is a necessary first step in many digital signal processing applications such as spectrum sensing, radar, and wireless communications. Analog-to-digital converters (ADCs) play a central role in most modern digital systems as they bridge the gap between the analog world and its digital counterpart. Yet, ADCs are a key implementation bottleneck in WSNs and many other \textit{Internet-of-Things (IoT)} applications due to the significant accrued power consumption and cost when they are used in large numbers.

In addition, the system bandwidth in a WSN is limited and thus is a fundamental constraint that must be considered. Hence, it is important to use a proper quantization scheme to reduce the communicated bits prior to transmission to address this limitations. Sampling at high data rates with high resolution ADCs would also dramatically increase the manufacturing cost of these electronic components. An immediate solution to such challenges is to use low-resolution, and specifically \textit{1-bit}, analog-to-digital converters (ADCs) \cite{kong2018nonlinear,jedda2017massive,mo2018mimo,stein2018one}. Therefore, the problem of recovering a signal from 1-bit measurements has attracted a great deal of interest over the past few years in a wide range of applications---see, e.g., \cite{jacobsson15,plan13, DEEPREC, jacques13, NAVEED,li2018bayesian,liu2018massive}, and the references therein. To name a few, the authors in \cite{masry1980reconstruction,cvetkovic2000single,ribeiro2006bandwidth,host2000effects,bar2002doa,dabeer2006signal,dabeer2008multivariate}, have investigated this problem from a classical statistical viewpoint. Particularly, 1-bit sampling and signal recovery have been extensively studied in the context of recently introduced one-bit Compressive Sensing (CS) problem as well \cite{zymnis2010compressed,plan2013one,jacques2013robust,yan2012robust,kamilov2012one,ai2014one,zhang2014efficient,boufounos20081,plan2013robust}. More specifically, the task of recovering the frequency and phase of temporal and spatial sinusoidal signals utilizing only 1-bit information with fixed quantization thresholds has been extensively investigated in \cite{host2000effects} and \cite{bar2002doa}, respectively. On the other hand, the recovery of general signals with high-dimensional parameters from sign comparison information were considered in \cite{dabeer2006signal} and \cite{dabeer2008multivariate}. In the context of CS, it was shown that sparse signals can be accurately recovered with high probability from 1-bit data when sufficient number of measurements is obtained \cite{plan2013one,jacques2013robust}. However, most of the CS literature have only considered the case of comparing the signal of interst with zero, which makes it impossible to recover the amplitude of the signal of interest.\looseness=-1
\par In this paper, we examine the most extreme case of quantization, i.e. 1-bit case, and propose an efficient signal estimation and threshold design algorithm which can perform the task of signal recovery from its 1-bit noisy measurements under the scenarios of white, colored or correlated noise. Furthermore, the proposed method can handle the task of parameter estimation for both cases of time-varying and fixed signals.\looseness=-1
\section{System Model}
\begin{figure*}[ht!]
	\centering
	\begin{subfigure}[b]{.48\linewidth}        
		\centering

		\centerline{\includegraphics[width=5.3cm]{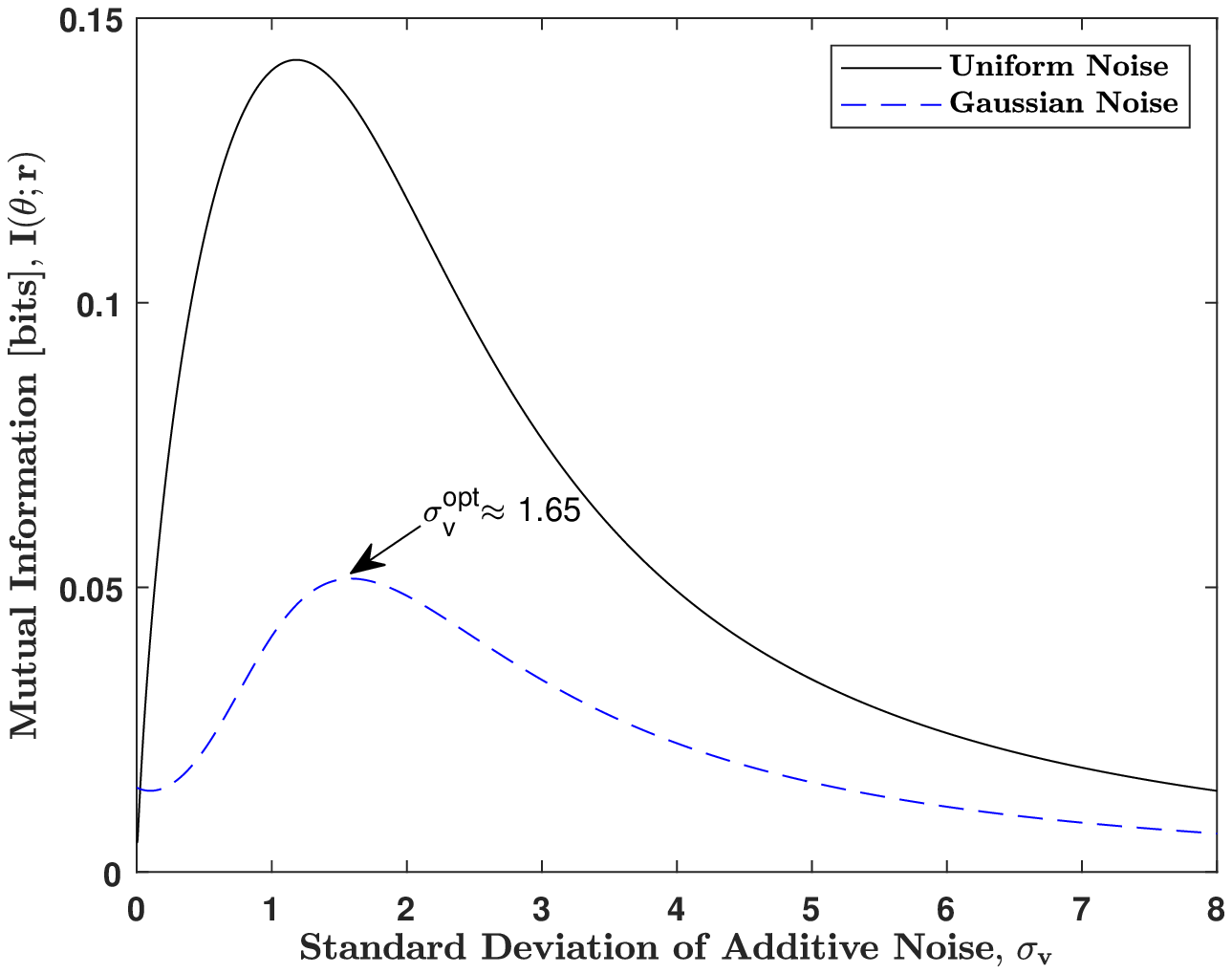}}
		\caption{}
		\label{fig:A}
	\end{subfigure}
	\begin{subfigure}[b]{.48\linewidth}        
		\centering

		\centerline{\includegraphics[width=5.3cm]{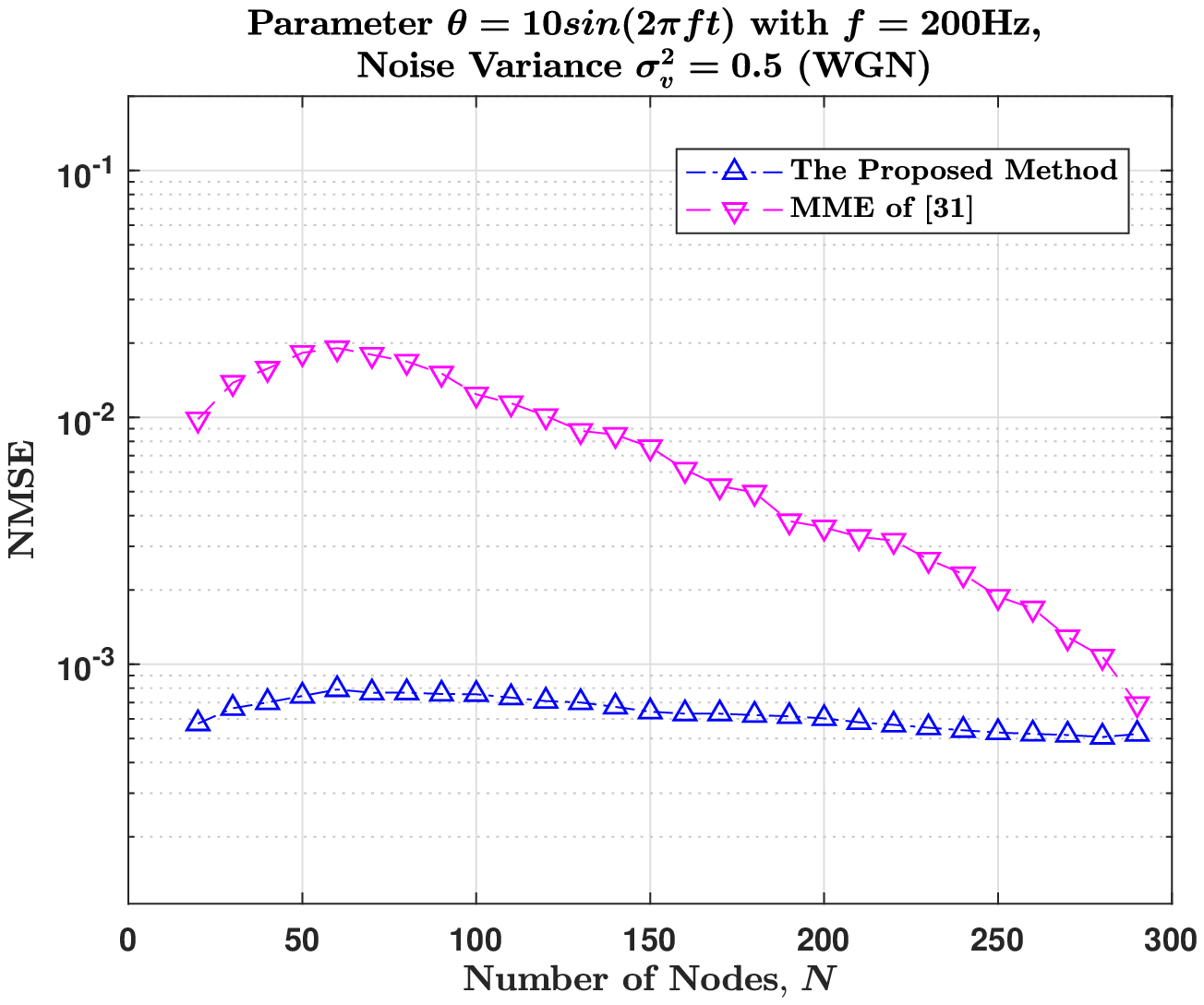}}
		\caption{}
		\label{fig:B}
	\end{subfigure}
	\caption{\small (a) demonstrates the mutual information $I(\theta;r)$ of the form \eqref{eq:15} for uniform and Gaussian input signal distributions, and (b) illustrates the performance of our proposed algorithm for a time-varying unknown parameter $\theta^{(k)}=10sin(2\pi fk)$, with $f=200\mbox{Hz}$.}
	\label{fig:roc_curve}
	\vspace{-3mm}
\end{figure*}
We consider a wireless sensor network with $N$ spatially distributed single-antenna nodes each of which observing an unknown deterministic parameter $\theta\in\mathds{R}$, at the time index $k$, according to the linear observation model of $z_i^{(k)} = \theta^{(k)} + v_i^{(k)}$,
where $i$ denotes the sensor index, and $v_i^{(k)}$ is additive zero-mean Gaussian observation noise with a Normal distribution, e.g., $v_i^{(k)}\sim\mathcal{N}(0,\sigma_v^2)$. Hence, the observed signal at all nodes can be compactly formulated as
\begin{equation}
\label{eq:2}
\mbf{z}^{(k)} = \theta^{(k)}\mathds{1} + \mbf{v}^{(k)},
\end{equation}
where $\mathds{1}$ denotes the all-one vector.

In order to satisfy the inherent bandwidth and power budget constraints in WSNs, we assume in this case that each node utilizes a \textit{1-bit quantization} scheme to encode its observation into 1 bit of information which will be transmitted to the fusion center for further processing. Namely, the $i$th node applies the following quantization function on its observed data $z_i^{(k)}$ prior to transmission:
\begin{equation}
\label{eq:3}
r_i^{(k)}=\text{sgn}\left(z_i^{(k)}-\tau_i^{(k)}\right),
\end{equation}
where $\mbox{sgn}(\cdot)$ is the \textit{sign} function, and $\tau_i^{(k)}$ denotes the quantization threshold at the $i$th node, for the time index $k$. Assuming that each node can reliably transmit 1 bit of information to the FC, the aggregated received data from all nodes at the $k$-th sampling time can be expresses as
\begin{equation}
\label{eq:4}
\boldsymbol{r}^{(k)} = \text{sgn}\left(\mbf{z}^{(k)}-\boldsymbol{\tau}^{(k)}\right)=\text{sgn}\left(\theta^{(k)}\mathds{1} + \mbf{v}^{(k)}-\bb{\tau}^{(k)}\right),
\end{equation}
where the sign function is applied element-wise, and $\mbf{v}^{(k)}~=~[v_1^{(k)},\dots,v_N^{(k)}]^{T}$ is the combined observation noise vector with covariance matrix $\boldsymbol{\Sigma}^{(k)}$, and $\bb{\tau}^{(k)}~=~[\tau_1^{(k)},\dots,\tau_N^{(k)}]$ is the quantization threshold vector. Next, the FC utilizes the received 1-bit information $\boldsymbol{r}^{(k)}$ to first construct an estimation of the unknown parameter $\theta^{(k)}$, and then, to further design the next quantization threshold for each node accordingly. 

An important observation which we use is that given a set of quantization thresholds $\boldsymbol{\tau}^{(k)}$, the corresponding vector of 1-bit measurements $\boldsymbol{r}^{(k)}$ defined in \eqref{eq:4} represents a limitation on the \emph{geometric} location of the unquantized data $\mbf{z}^{(k)}$. Particularly, one can capture this geometric knowledge on $\mbf{z}^{(k)}$ through the following linear inequality:
\begin{equation}
\label{eq:5}
\bb{\Omega}^{(k)}\left(\mbf{z}^{(k)}-\bb{\tau}^{(k)}\right)\boldsymbol\succeq\mbf{0},
\end{equation}
where $\bb{\Omega}^{(k)}=\text{\text{Diag}}\{\mbf{\boldsymbol{r}^{(k)}}\}$, and $\text{Diag}\{\cdot\}$ is the diagonalization operator, and $\mbf{0}$ denotes the all-zero vector.
\section{Centralized 1-Bit Signal Recovery via Quadratic Programming}
\begin{figure*}[ht!]
	\centering
	\begin{subfigure}[b]{.48\linewidth}        
		\centering
		\centerline{\includegraphics[width=5.3cm]{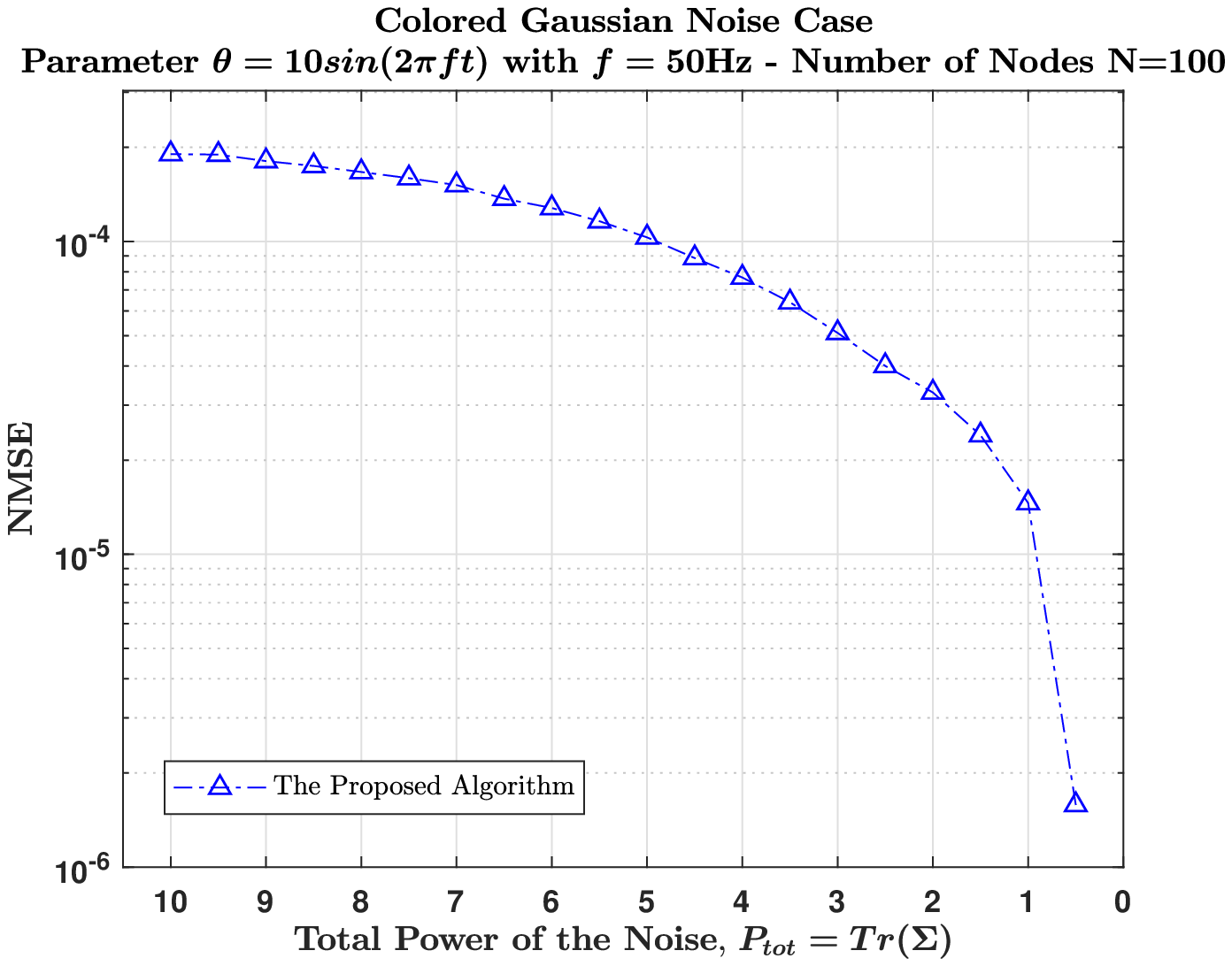}}
		\caption{}
		\label{fig:A}
	\end{subfigure}
	\begin{subfigure}[b]{.48\linewidth}        
		\centering
		\centerline{\includegraphics[width=5.3cm]{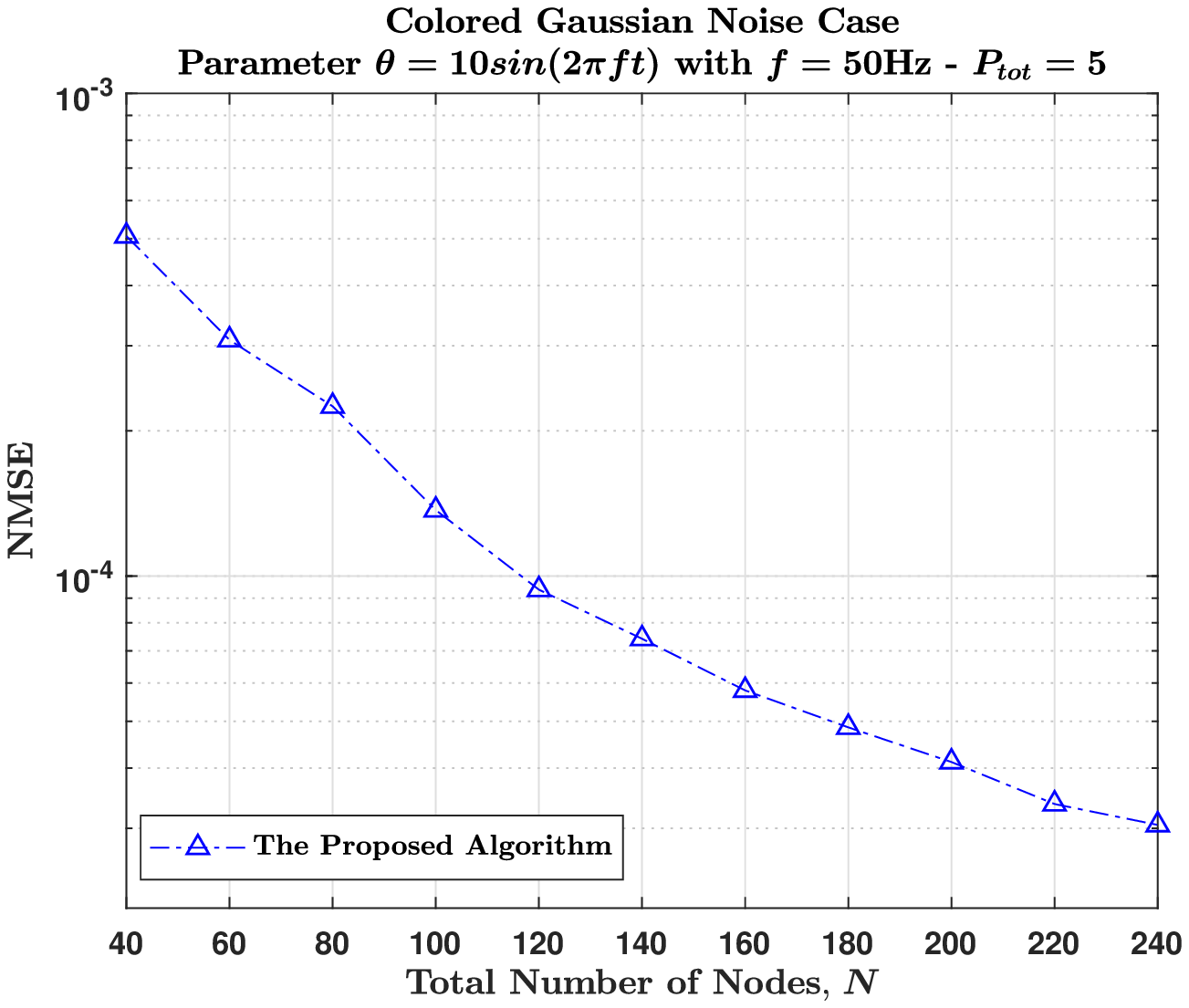}}
		\caption{}
		\label{fig:B}
	\end{subfigure}
	\caption{\small (a) illustrates the NMSE versus the total power of the \emph{colored} Gaussian noise $P_{tot}=Tr(\bb{\Sigma})$ for a network with $N=100$, when the unknown parameter $\theta^{(k)}=10sin(2\pi ft)$, with $f=50\mbox{Hz}$, and (b) shows the NMSE versus the total number of nodes $N$, for the same signal and noise model.}
	\label{fig:roc_curve}
\end{figure*}
We lay the ground for our 1-bit statistical and inference model using the weighted least square (WLS) method. Let $\boldsymbol{\Sigma}=\mathds{E}\{\mbf{v}^{(k)}\mbf{v}^{(k)^H}\}$ denote the covariance matrix of the noise vector $\mbf{v}^{(k)}$. Note that if the unquantized information vector $\mbf{z}^{(k)}$ was available at the FC, then the maximum likelihood (ML) estimation of the unknown parameter $\theta^{(k)}$ given $\mbf{z}^{(k)}$ can be expressed as:
\begingroup
\setlength{\abovedisplayskip}{7pt}
\setlength{\belowdisplayskip}{7pt}
\begin{equation}
\label{eq:6}
\hat{\theta}^{(k)} = [\mathds{1}^{T}\bb{\Sigma}^{-1}\mathds{1}]^{-1}\mathds{1}^{T}\bb{\Sigma}^{-1}\mbf{z}^{(k)},
\end{equation}
\endgroup
where $\hat{\theta}^{(k)}$ denotes the ML estimate of the unknown parameter according to the vector of observations $\mbf{z}^{(k)}$. Furthermore, it is well-known that the variance of the ML estimation in \eqref{eq:6} is given by $\text{Var}(\hat{\theta}^{(k)})=[\mathds{1}^{T}\bb{\Sigma}^{-1}\mathds{1}]^{-1}$. Alternatively, one can obtain the maximum-likelihood estimator of the unknown parameter by minimizing the following WLS criterion,
\begin{align}
\label{eq:7}
\mathcal{Q}(\mbf{z},\theta) := ||\mbf{z} - \mathds{1}\theta||_{\boldsymbol\Sigma^{-1}}^2=\left(\mbf{z} - \mathds{1}\theta\right)^{H}\boldsymbol{\Sigma}^{-1}\left(\mbf{z} - \mathds{1}\theta\right),
\end{align}
where $\mathcal{Q}(\mbf{z},\theta)$ is our cost function to be minimized over the parameters $(\mbf{z},\theta)$. A natural approach to obtain an estimate of $\theta^{(k)}$ using the 1-bit quantization vector $\boldsymbol{r}^{(k)}$, is to use an alternating optimization approach and further exploit the limitation on the geometric location of $\mbf{z}^{(k)}$ imposed by \eqref{eq:5}; in other words, to first obtain an estimate of $\mbf{z}^{(k)}$ by fixing the variable $\theta^{(k)}$ and then to recover the unknown parameter $\theta^{(k)}$ using \eqref{eq:6}, given by $\text{\emph{argmin}}_{\theta}\mathcal{Q}(\mbf{z},\theta)$. Interestingly, for a fixed parameter $\mbf{z}$, the optimal $\theta^{*}$ that minimizes $\mathcal{Q}(\mbf{z},\theta)$ coincides with that of the MLE of $\theta^{(k)}$ given in \eqref{eq:6}.

We can further substitute the optimal $\hat{\theta}^{(k)}$ of \eqref{eq:6} into \eqref{eq:7} to simplify the objective function $\mathcal{Q}(\mbf{z},\theta)$ in terms of the parameter $\mbf{z}$, viz.
\begin{align}
\label{eq:8}
\mathcal{Q}(\mbf{z}) = \mathcal{Q}(\mbf{z},\hat{\theta}^{(k)}) = \mbf{z}^H(\mbf{I}-\mathds{1}\boldsymbol\eta^H)^H\boldsymbol{\Sigma}^{-1}(\mbf{I}-\mathds{1}\boldsymbol\eta^H)\mbf{z},
\end{align}
where we define $\boldsymbol{\eta}$ as,
\begin{align}
\label{eq:9}
\boldsymbol{\eta}\triangleq\frac{\boldsymbol{\Sigma}^{-1}\mathds{1}}{\mathds{1}^T\boldsymbol{\Sigma}^{-1}\mathds{1}}.
\end{align}
Consequently, one can cast the problem of recovering the unquantized vector $\mbf{z}$ from 1-bit noisy measurements $\boldsymbol{r}^{(k)}$ as the following constrained quadratic program (CQP):
\begin{align}
\label{eq:10}
\hat{\mbf{z}}^{(k)} =\; &\underset{\mbf{z}}{\text{\emph{argmin}}}\qquad
\mbf{z}^H(\mbf{I}-\mathds{1}\boldsymbol\eta^H)^H\boldsymbol{\Sigma}^{-1}(\mbf{I}-\mathds{1}\boldsymbol\eta^H)\mbf{z}, \\
\label{eq:11}
& \text{\emph{subject to}}\;\;\;
\boldsymbol\Omega^{(k)}(\mbf{z}-\boldsymbol\tau^{(k)})	\boldsymbol\succeq \mbf{0},
\end{align}
where the inequality in \eqref{eq:11} is applied element-wise (equivalent to $N$ scalar inequality constraints). Note that the constraint \eqref{eq:11} ensures the consistency between the received 1-bit quantized data $\boldsymbol{r}^{(k)}$ (as incorporated in $\boldsymbol{\Omega}^{(k)}$) and the solution $\hat{\mbf{z}}^{(k)}$. Moreover, note that the matrix $\mbf{M}~=~(\mbf{I}~-~\mathds{1}\boldsymbol\eta^H)^H\boldsymbol{\Sigma}^{-1}(\mbf{I}~-~\mathds{1}\boldsymbol\eta^H)$ at the core of this CQP is positive semi-definite. Therefore, the CQP problem in \eqref{eq:10} is convex and can be solved efficiently using standard numerical methods (e.g., the interior point method \cite{boyd04convex}).

Having obtained $\hat{\mbf{z}}^{(k)}$ via solving \eqref{eq:10}, we can then easily estimate the unknown parameter $\theta^{(k)}$ by solving the program,
\begin{equation}
\label{eq:26}
\underset{\theta}{\text{argmin }}\mathcal{Q}(\hat{\mbf{z}}^{(k)},\theta),
\end{equation}
whose closed-form solution $\hat{\theta}^{(k)}$ is given in \eqref{eq:6}.
\section{Quantization Threshold Design}
So far, we discussed our inference framework to estimate an unknown deterministic signal from its 1-bit noisy measurements and the corresponding quantization thresholds. In order to further facilitate the estimation process and to restore the exact signal model (both amplitude and phase), it is of essence to design a proper adaptive quantization thresholding scheme. In this section, we devise a stochastic adaptive thresholding method for our estimation algorithm, which thanks to its adaptive nature, empowers the proposed inference framework to accurately recover time-varying signals as well. We first investigate the performance of a 1-bit quantizer of the form \eqref{eq:3} in the presence of additive noise from an information theoretic viewpoint, and further show that the presence of noise is indeed improving the performance of a 1-bit quantization scheme. We then propose our threshold design strategy accordingly.\looseness=-1

\par We first consider the case that the unknown parameter is a random variable with a known distribution, and that the observation noise $v\sim\mathcal{N}(0,\sigma_v^2)$. Clearly, each 1-bit sample $r=\text{sgn}(\theta+v-\tau)$, (at a given time index and quantization threshold value $\tau$) can be seen as a random variable that follows a Bernoulli distribution $\mathcal{B}(p_{\theta})$, whose parameter $p_{\theta}$ is given by
\begingroup
\setlength{\abovedisplayskip}{4pt}
\setlength{\belowdisplayskip}{4pt}
\begin{equation}
\label{eq:14}
p_\theta \triangleq P_{r|\theta}(1|\theta) = \text{Pr}(\theta+v_i>\tau_i) = Q\left(\frac{\tau_i-\theta}{\sigma_v}\right),
\end{equation}
\endgroup
where $P_{r|\theta}$ represents the conditional probability of receiving $r=1$ given the parameter $\theta$, and $Q(\cdot)$ denotes the standard Q-function. The mutual information \cite{cover2012elements} between the unknown parameter $\theta$ and the obtained 1-bit sample $r_i$ can be expressed as
\begin{align}
\label{eq:15}
I(\theta;r) &= \mathcal{H}(r) - \mathcal{H}(r|\theta)\\
			&=-\sum_{i=0}^{1}P_r(i)\text{log}_2P_r(i)
			-\int_{\mathds{R}} f_{\theta}(\theta^\prime)\mathcal{H}(r|\theta=\theta^\prime)d\theta^\prime\nonumber,
\end{align}
where $P_r(i)$ for $i\in\{0,1\}$, is the probability mass function (pmf) of the discrete Bernoulli random variable $r$, and $f_\theta$ is the probability density function (pdf) of the parameter of interest $\theta$, and $\mathcal{H}(\cdot)$ denotes the entropy function of the argument random variable. Moreover, the conditional entropy $\mathcal{H}(r|\theta=\theta^\prime)$ in \eqref{eq:15} can be further simplified in terms of the noise distribution and according to \eqref{eq:14} as follows:
\begin{equation}
\label{eq:16}
	\mathcal{H}(r|\theta=\theta^\prime) = p_{\theta^{\prime}}\text{log}_2p_{\theta^{\prime}} + (1-p_{\theta^{\prime}})\text{log}_2(1-p_{\theta^{\prime}}).
\end{equation}
Moreover, the pmf of the Bernoulli random variable $r$ can be recast as
\begin{equation}
\label{eq:17}
P_r(i) = \int_{\mathds{R}}f_{\theta}(\theta^{\prime})P_{r|\theta}(i|\theta=\theta^{\prime})d\theta^{\prime}.
\end{equation}
Eventually, one can easily calculate the mutual information of the form \eqref{eq:15} between the unknown parameter and the observed 1-bit samples by utilizing \eqref{eq:14}-\eqref{eq:17}. 
\par Figure 1 (a) illustrates the mutual information $I(\theta;r)$ versus the Gaussian noise standard deviation $\sigma_v$ for two cases: (1) $\theta$ is uniformly distributed and (2) $\theta$ follows a Normal distribution. Surprisingly, as the noise power increases, the mutual information shows a non-monotonically behavior in the presence of noise in both cases. Namely, the mutual information between the input signal, and the output of the 1-bit quantizer $r$, first achieves a global maximum and then dampens. This implies that a moderate amount of noise can indeed provide signal processing benefits in the case of 1-bit quantization schemes. Henceforth, this motivates us to employ a stochastic threshold design method in which we artificially induce noise to the system through the thresholds $\boldsymbol{\tau}^{(k)}$, to exploit this non-monotonic behavior of mutual information $I(\theta;r)$. In addition, we further consider the current knowledge of the unknown parameter at each time index to tune the quantization thresholds of each node for the next observation period, which enables us to achieve an even more accurate estimate of the parameter of interest.

\textbf{--- Stochastic Threshold Design:}  Herein, we propose our adaptive threshold design strategy for the task of 1-bit signal recovery. It can be shown that the optimal threshold given the Bernoulli observations $\boldsymbol{r}^{(k)}$ is indeed equal to the unknown parameter; i.e., $\tau^{(k)}_{opt}=\theta^{(k)}$ at each time index. Nevertheless, we cannot use the optimal threshold in that it is a function of the unknown parameter at each sampling period, and therefore, cannot be used in practice. Hence, a natural approach to determine the next quantization threshold is to exploit our current knowledge of the unknown parameter to set the next quantization thresholds. Namely, the fusion center should first obtain an estimate of the parameter based on the received quantized information $\boldsymbol{r}^{(k)}$ from the nodes, and then set the next quantization threshold for each node according to the obtained estimate of the unknown parameter (note that $\hat{\mbf{\theta}}^{(k)}$ is our best estimate of the real value of the unknown parameter $\theta^{(k)}$, at the time index $k$). With the current estimate of the unknown parameter at hand, the FC samples the next quantization threshold for each node from a Normal distribution with mean $\hat{\theta}^{(k)}$, and variance $\sigma_{\tau}^2$. In other words, the fusion center adds $N$ realizations of a zero-mean random variable $w_{\tau}\sim\mathcal{N}(0,\sigma_{\tau}^2)$ to the current estimate of the unknown parameter obtained from \eqref{eq:26} in order to choose the next thresholds. Namely, after obtaining $\hat{\theta}^{(k)}$ at the $k$th cycle, the FC chooses the next quantization thresholds $\boldsymbol{\tau}^{(k+1)}$ according to the following model,

\begin{equation}
\bb{\tau}^{(k+1)} = \hat{\theta}^{(k)}\mathds{1} + \mbf{w}^{(k)}_{\tau},
\end{equation} 
where $\mbf{w}_{\tau}^{(k)}=[w_{1}^{(k)},\dots,w_{N}^{(k)}]^T$ are $N$ independent samples drawn from the normal distribution of $\mathcal{N}(0,\sigma_{\tau}^2)$. Note that, by using a stochastic threshold design strategy, we are able to introduce an artificial noise whose variance can be controlled in such a way to not only maximize the mutual information but also incorporate the current estimate of the unknown parameter in the design. Furthermore, the variance of the random variable $w_{\tau}$ can be chosen according to the observation noise variance to maximize the mutual information. More generally, we can model $\bb{\tau}^{(k+1)}$ as a multivariate Gaussian random vector with mean vector $\bb{\mu}^{(k+1)}=\theta^{(k)}\mathds{1}$, and covariance matrix $\bb{\Sigma}_{\tau}$, e.g., $\bb{\tau}^{(k+1)}\sim\mathcal{N}(\bb{\mu}^{(k+1)},\bb{\Sigma}_{\tau})$.

The proposed signal recovery and threshold design method is summarized in Table~1.
\begin{table}[t]
	\vspace{-.2cm}
	\begin{algorithm}[H]
		\small
		\caption{The Proposed Adaptive Signal Recovery Method.}\label{t:1}
		\textbf{Step 0:} Initialize the thresholds vector $\bb{\tau}^{(0)}\in\mbf{R}^N$.\\
		\textbf{Step 1:} Each node performs the quantized measurement of the form $r_i^{(k)}=\text{sgn}(z_i^{(k)}-\tau_i^{(k)})$, and transmits $r_i^{(k)}$ to the FC ($i$ denotes the node index).\\
		\textbf{Step 2:} FC constructs the quantized matrix $\boldsymbol{\Omega}^{(k)} = \text{Diag}\{\boldsymbol{r}^{(k)}\}$, based on the received 1-bit measurements vector $\boldsymbol{r}^{(k)}$, and recover $\hat{\mbf{z}}^{(k)}$ via solving the proposed CQP in \eqref{eq:10}. Namely,
		\begin{enumerate}
			\item[] 
			$\hat{\mbf{z}}^{(k)} = \text{argmin}_\mbf{z}\,\,\mathcal{Q}(\mbf{z})\;\;\;\text{s.t. }\boldsymbol\Omega^{(k)}(\mbf{z}-\boldsymbol\tau^{(k)})\succeq\mbf{0}.$
		\end{enumerate}
		\textbf{Step 3:} Given $\mbf{\hat{z}}^{(i)}$ from the previous step, FC estimates the unknown parameter $\theta^{(k)}$ the ML/WLS estimate given in \eqref{eq:6}.\\
		\textbf{Step 4:}
		The fusion center chooses the next quantization threshold for each node according to the following model,
		\begin{enumerate}
			\item[] $\boldsymbol\tau^{(k+1)}=\hat{\theta}^{(k)}\mathds{1}+\mbf{w}_{\tau}^{(k)}$.
		\end{enumerate}
		where $\mbf{w}_{\tau}^{(k)}=[w_1^{(k)},\dots,w_N^{(k)}]^T$ denotes $N$ realizations of the random variable $w_\tau\sim\mathcal{N}(0,\sigma_{\tau}^2)$. Repeat steps 1-4 until convergence.
	\end{algorithm}
	\vspace{-9mm}
\end{table}
\section{Numerical Results}
In this section, we evaluate the performance of the two proposed algorithms for the task of 1-bit signal recovery. 
We define the normalized mean square error (NMSE) of an estimate $\hat{\mbf{x}}$ of a signal $\mbf{x}$ as
\begin{equation}
\text{NMSE}\triangleq\mathds{E}\left\{\frac{||\mbf{x}-\hat{\mbf{x}}||^2_2}{||\mbf{x}||^2_2}\right\}.
\end{equation}
Each data point presented in the numerical results is averaged over $100$ independent samples and realizations of the problem parameters. In the following, we analyze the performance of our proposed algorithms in different scenarios. In addition, we compare the performance of our method in the presence of white Gaussian noise (WGN) with the modified mean estimator (MME) proposed in \cite{wu09}. It must be noted that our algorithm can handle the task of parameter estimation in the presence of both white or colored (correlated) noise. However, the MME method of \cite{wu09} can only handle the scenario of white Gaussian noise.
\par Fig.  1(b) demonstrates the normalized mean square error (NMSE) vs. the total number of nodes $N$, for the white Gaussian noise (WGN) scenario when $\theta^{(k)}=10sin(2\pi fk)$ where $f=200\mbox{Hz}$. It can be seen that in the presence of WGN, our proposed method significantly outperforms the MME method of \cite{wu09} and attains a very high accuracy for estimating the unknown parameter. 
\par In Fig. 2 we consider the presence of (correlated) colored Gaussian noise at the time of observation, where each node observes a time-varying parameter of the form $\theta^{(k)}=10sin(2\pi fk)$ with $f=50\mbox{Hz}$. Particularly, Fig. 2(a) illustrates the performance of our proposed method versus the total power of colored noise $P_{tot}=\text{Tr}(\bb{\Sigma})$, and it can be seen that in the presence of colored noise, the proposed method can accurately recover the unknown time-varying parameter. On the other hand, Fig. 2(b) demonstrates the NMSE vs. the total number of nodes assuming that $P_{tot}=5$. It can be seen that as the number of nodes $N$ (1-bit information) increases, the accuracy of our proposed method improves, and in most of the cases, the NMSE  attains values that are virtually zero (note that we used logarithmic scale in our illustrations).
\section{Conclusion}
In this paper, we assumed the most extreme case of quantization, i.e. the 1-bit case, and proposed an efficient signal recovery and threshold design method which can perform the task of signal recovery from its 1-bit noisy measurements under both scenarios of the presence of white and colored Gaussian noise. Moreover, the proposed algorithms can accurately recover fixed as well as time-varying unknown parameters (e.g., a sinusoidal signal).
\nocite{}
\bibliographystyle{IEEEtran}
\balance
\bibliography{IEEEabrv,refs}
\end{document}